\documentclass[runningheads]{llncs}

\usepackage{subfigure}

\usepackage{graphicx}
\usepackage{xr-hyper}
\usepackage{hyperref}
\usepackage{multirow}
\usepackage{url}            
\usepackage{booktabs}       
\usepackage{amsfonts}       
\usepackage{nicefrac}       
\usepackage{microtype}      
\usepackage{xcolor}
\usepackage{graphicx}
\usepackage{amsmath}
\usepackage{amssymb}
\usepackage{authblk}

\newcommand{\myparagraph}[1]{\smallskip\noindent\textbf{#1}}

\begin{document}
\title{A Topology-Attention ConvLSTM Network and Its Application to EM Images}
\titlerunning{Topology-Attention ConvLSTM Network}

\author{Jiaqi Yang\inst{1\star}\and
Xiaoling Hu\inst{2\star} \and
Chao Chen\inst{2} \and
Chialing Tsai\inst{1} }

\renewcommand{\thefootnote}{\fnsymbol{footnote}}
\footnotetext[1]{The two authors contributed equally to this paper.}

\institute{Graduate Center, CUNY \and Stony Brook University}

\maketitle             

\begin{abstract}
Structural accuracy of segmentation is important for fine-scale structures in biomedical images. We propose a novel Topology-Attention ConvLSTM Network (TACNet) for 3D image segmentation in order to achieve high structural accuracy for 3D segmentation tasks.
Specifically, we propose a Spatial Topology-Attention (STA) module to process a 3D image as a stack of 2D image slices and adopt ConvLSTM to leverage contextual structure information from adjacent slices. In order to effectively transfer topology-critical information across slices, we propose an Iterative-Topology Attention (ITA) module that provides a more stable topology-critical map for segmentation. Quantitative and qualitative results show that our proposed method outperforms various baselines in terms of topology-aware evaluation metrics.
\end{abstract}

\section{Introduction}

\noindent

\begin{figure}[t]
\centering
\subfigure{
\includegraphics[width=0.15\textwidth]{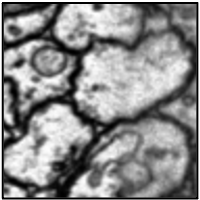}}
\hspace{-.08in}
\subfigure{
\includegraphics[width=0.148\textwidth]{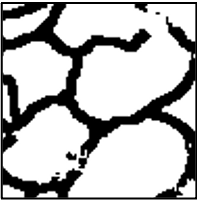}}
\hspace{-.08in}
\subfigure{
\includegraphics[width=0.149\textwidth]{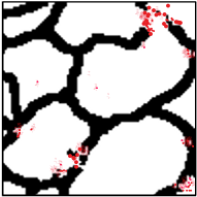}}
\hspace{-.08in}
\subfigure{
\includegraphics[width=0.15\textwidth]{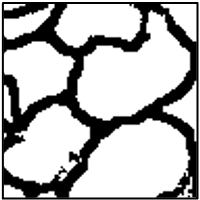}}
\hspace{-.08in}
\subfigure{
\includegraphics[width=0.148\textwidth]{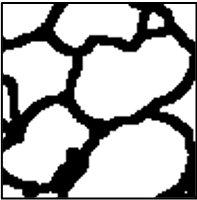}}
\caption{From left to right: original image, ConvLSTM result, generated attention map across adjacent slices (overlay with segmentation), our result, and ground truth.}.
\label{teaser}
\end{figure}

Deep learning methods have achieved state-of-the-art performance for image segmentation.
However, most existing methods focus on per-pixel accuracy (e.g., minimizing the cross-entropy loss) and are prone to structural errors, e.g., missing connected components and broken connections. These structural errors can be fatal in downstream analysis, affecting the functionality of the extracted fine-scale structures such as neuron membranes, vessels and cells. See Fig.~\ref{teaser}.

To address this issue, differentiable topological losses~\cite{hu2019topology,clough2019explicit,hu2021topologyaware,yang20213d} have been proposed to enforce the network to learn to segment with correct topology. However, these methods have their limitations when applied to 3D images, due to the high computational cost of topological information.
Furthermore, we often encounter anisotropic images, but the topological loss cannot be applied for 3D anisotropic images directly. For example, a tube in 3D may manifest as a series of rings across different slices rather than a seamless tube. The topology of the tube cannot be captured even in the ground truth annotation. 

In this paper, a novel 3D topology-preserving segmentation method is proposed to address the aforementioned issues.
Inspired by existing approaches for anisotropic images \cite{nunez2013machine}, we propose to first segment individual slices, and then stack the segmentation results together as the 3D output. 
This way, the topological computation is restricted within each 2D slice, and thus can be very efficient.
To achieve a topology-preserving segmentation model in this slice-by-slice approach, the key challenge is how to effectively share the topological information between adjacent slices. 
A successful method should account for the fact that the topology of consecutive slices share some similarity, but are not the same. When segmenting one slice, the topology of other slices should help recalibrate the prediction, but in a soft manner.
We use \emph{convolutional LSTM (ConvLSTM)}~\cite{chen2016combining} as our backbone. Specifically designed for 3D anisotropic images, ConvLSTM uses 2D convolution and exploits inter-slice correlation to achieve high quality results while being more efficient than 3D convolutional nets. 
To propagate topological information across slices, we propose a \emph{Spatial Topology-Attention module}, which redirects the convolutional network's attention toward topologically critical locations of each slice, based on the topology of itself and its adjacent slices.

Another challenge is that the topologically critical map can be inconsistent across different slices and unstable through training epochs. This is mainly because the topological information is extracted from each 2D slice rather than from 3D. 
To this end, furthermore, we propose an \emph{Iterative Topological-Attention module} that iteratively refines the critical map through epoches. 

Our method, called \emph{Topology-Attention ConvLSTM Network (TACNet)}, fully utilizes topological information from adjacent slices for 3D images without much additional computational cost. Empirically, our method outperforms baselines in terms of topology-aware metrics. In summary, our main contribution is threefold: 
\begin{enumerate}
    \item A novel Spatial Topology-Attention (STA) module to capture spatial contextual topological information from adjacent slices.
    \item An Iterative Topology-Attention (ITA) module to improve the stability of the critical points, and consequently the quality of the final results.
    \item Combining Topology-Attention with ConvLSTM to achieve high performance on 3D image segmentation benchmarks.
\end{enumerate}

\section{Related Works}

Standard 3D medical image segmentation methods directly apply the networks to 3D images~\cite{cciccek20163d,milletari2016v,meirovitch2019cross,funke2017deep}. These methods could be computationally expensive. 
Alternatively, one may first segment each 2D slice, and then link the 2D segmentation results to generate 3D results~\cite{nunez2013machine,yang20213d}. 
Note that this segment-then-link approach ignores the contextual information shared among adjacent slices at the segmentation step. To address this, one may introduce pooling techniques across adjacent slices~\cite{fakhry2016deep}. But these methods are not explicitly modeling the topology as our method does.

\myparagraph{Persistent homology.}
Our topological approach is based on the theory of persistent homology~\cite{edelsbrunner2010computational,edelsbrunner2000topological}, which has attracted a great amount of attention both from theory~\cite{bubenik2015statistical,fasy2014confidence} and from applications~\cite{kulp2015ventricular,wong2016kernel}. 
In image segmentation, 
persistent-homology-based topological loss functions~\cite{hu2019topology,clough2019explicit} have been proposed to train a neural network to preserve the topology of the segmentation. 
The key insight of these methods is to identify critical locations for topological correctness, and improve the neural network's prediction at these locations.
These critical locations are computed using the theory of persistent homology, and correspond to critical points (local maxima/minima and saddles) of the likelihood function. 

\myparagraph{Attention mechanism.}
Attention modules model relationships between pixels/channels/feature maps and have been widely applied in both vision and natural language processing tasks~\cite{lin2016efficient,lin2017structured,shen2017disan}. Specifically,  self-attention mechanism~\cite{vaswani2017attention} is proposed to draw global dependencies of inputs and has been used in machine translation tasks. \cite{zhang2019self} tries to learn a better image generator via self-attention mechanism. \cite{wang2018non} mainly explores effectiveness of non-local operation, which is similar to self-attention mechanism. \cite{zhao2018psanet} learns an attention map to aggregate contextual information for each individual point for scene parsing.

\section{Method}

\begin{figure*}[t]
\centering
\includegraphics[width=0.75\textwidth]{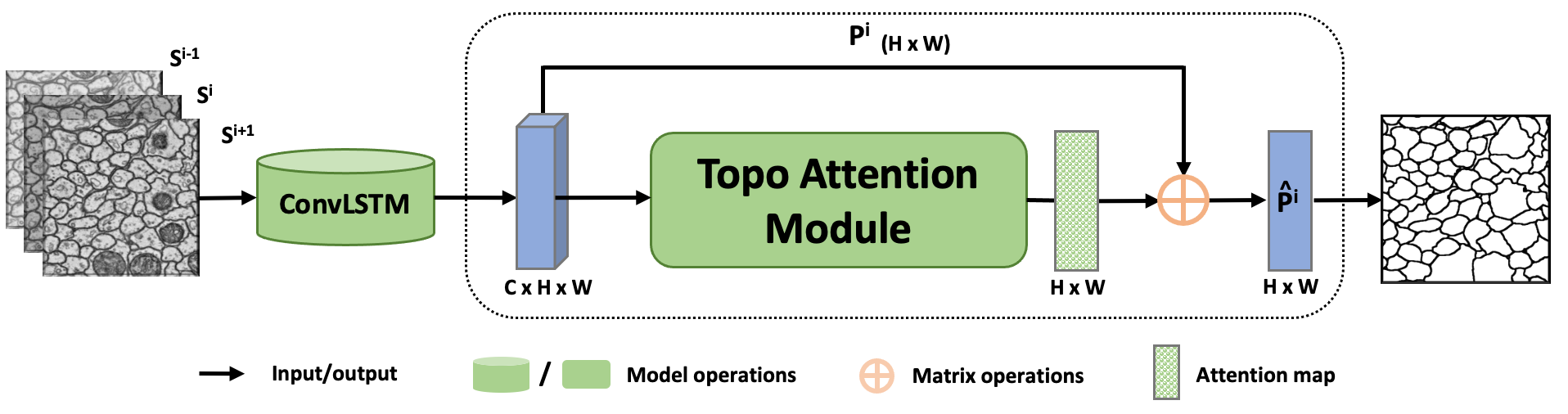}
\caption{Overview of the proposed framework}.
\label{overview}
\end{figure*}

The overview of the proposed architecture is illustrated in Fig.~\ref{overview}. To capture the inter-slice information, $l$ consecutive slices along Z-dimension are fed into a ConvLSTM. We set $l=3$ when describing our method for simplicity.
ConvLSTM is an extension of FC-LSTM, which has the convolutional operators in LSTM gates and is particularly efficient in exploiting image sequences. 
Note that the inputs to ConvLSTM are three adjacent slices, $\{S^{i-1}, S^i, S^{i+1}\} \in R^{H \times W}$, and the output also has three channels, $\{P^{i-1}, P^i, P^{i+1}\} \in R^{H \times W}$, each being the probabilistic map $P^i$ of the corresponding input slice $S^i$. We use $i-1, i, i+1$ to represent input slice indices in this paper. 

The three probabilistic maps $\{P^{i-1}, P^i, P^{i+1}\}$ are then fed into the Topology-Attention module. In this module, each pixel in the feature maps gathers rich structural information from both the current and adjacent slices, without introducing extra parameters. 
We propose a Spatial Topology-Attention module to model the correlation between the topologically critical information of adjacent slices. A Topology-Attention map is generated to highlight the locations which are vital for structural accuracy by taking topological information of adjacent slices into consideration. The details are further explained in Sec.~\ref{sec:inter}.

One challenge in using the topologically critical maps from different slides is that they are not stable and consistent across slides and across training epochs. 
Specifically, during the training process, the predicted probability maps will change slightly, while the corresponding Topology-Attention maps can be quite different, leading to instability of the training process. To further improve the stability of the engine, we propose an Iterative Topology-Attention, to adjust the attention maps from earlier epochs. The details can be found in Sec.~\ref{sec:iter}.

\subsection{Spatial Topology-Attention (STA) Module}
\label{sec:inter}

Continuation in contextual information across slices is essential for 3D image understanding, which can be obtained by taking adjacent slices into consideration. In order to collect contextual information in the Z-dimension to enhance the prediction quality, we introduce a STA module which encodes the inter-slices contextual information into the focused slice. 

\begin{figure}[t]
\centering
\includegraphics[width=0.7\columnwidth]{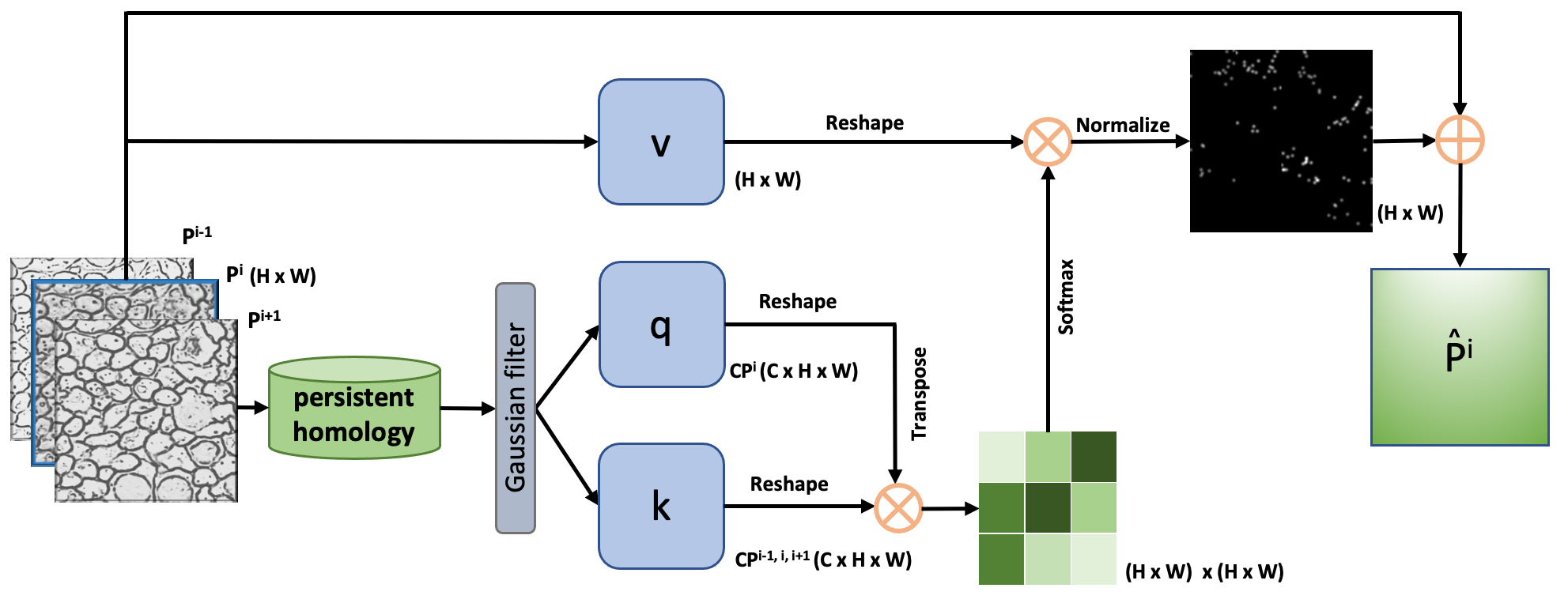} 
\caption{Illustration of the Spatial Topology-Attention (STA) module}
\label{spatial}

\end{figure}

As illustrated in Fig.~\ref{overview}, we can obtain three predicted probabilistic maps, $\{P^{i-1}, P^{i}$, $P^{i+1}\} \in R^{H \times W}$ which corresponds to the input slices after ConvLSTM. Fig.~\ref{spatial} shows the complete process that passes the probabilistic maps to the STA module, and yields the final probabilistic map $\hat{P^i}$ in the end. Next, we elaborate the process to aggregate topological contexts of inter-slices.

\myparagraph{Persistent homology and critical points.}
Given a 2D image likelihood map, we obtain the binary segmentation by thresholding at $\alpha = 0.5$. The 2D likelihood map can be represented as a 2D continuous-valued function $f$. We consider thresholding the continuous function $f$ with all possible thresholds. For a specific threshold $\alpha$, we define the thresholded results $ f^{\alpha}:= \{x \in \Omega | f(x) \geq \alpha\}$. By decreasing $\alpha$, we obtain a monotonically growing sequence $\varnothing \subseteq f^{\alpha_1} \subseteq f^{\alpha_2} \subseteq ... \subseteq f^{\alpha_n} = \Omega$, where $\alpha_1 \geq \alpha_2 \geq ... \geq \alpha_n$. As $\alpha$ changes, the topology of $f^{\alpha}$ changes. Consequently, some new topological structures are born while existing ones are killed. The theory of persistent homology captures all the birth time and death time of these topological structures and summarize them as a \emph{persistence diagram}. One can define a topological loss as the matching distance between the persistence diagrams of the likelihood function and the ground truth. When the loss is minimized, the two diagrams are the same and the likelihood map will generate a segmentation with the correct topology. 

As shown in \cite{hu2019topology}, the topological loss can be written as a polynomial function of the likelihood function at different critical pixels. These critical pixels correspond to critical points of the likelihood function (e.g., saddles, local minima and local maxima), and these critical points are crucial locations at which the current model is prone to make topological mistakes. The loss essentially forces the network to improve its prediction at these critical locations.

The likelihood maps at different slices generate different critical point maps. 
Here we propose an attention mechanism to aggregate these critical point maps across different slices to generate topology attention map for the current slice. 
The third column in Fig.~\ref{teaser} illustrate the generated attention map (overlay with segmentation) across adjacent slices.
For $\{P^{i-1}, P^{i}$, $P^{i+1}\}$, by using persistent homology algorithm, we identify the critical points. Here we use a Gaussian operation to expand the isolated critical points to a region because the surrounding regions will also be vital for structures. So we can obtain a soft version of critical point map: $CP^{i} = Gaussian(PH(P^{i}))$.
Here, $PH(\cdot)$ is the operation to generate isolated critical points and $Gaussian(\cdot)$ is a Gaussian operation. $CP^{i}$ has the same dimension as $P^{i} \in R^{H \times W}$.

To improve the computational efficiency, we combine the $\{CP^{i-1}, CP^{i}, CP^{i+1} \}$ into one single $k \in R^{C \times H \times W}$ $(C = 3)$ and expand the $CP^{i}$ into same size as $q \in R^{C \times H \times W}$. To obtain the correlation between target map ($q$) and consecutive slices ($k$), we reshape them to $R^{C \times N}$, where $N=H \times W$, and perform a matrix multiplication between the transpose of $q$ and $k$. The similarity map $SM \in R^{N \times N}$ is generated after a softmax. Formally,

\begin{equation}
\label{atten}
    SM_{nm} = \frac{exp(r_s(q_m, k_n))}{\sum_{m=1}^{N} exp(r_s(q_m, k_n))}
\end{equation}
where $r_s(q_m, k_n)$ computes the relation (eg. similarity) between $q_m$ and $k_n$. $SM_{nm}$ measures how strong correlation between two pixels is.

Next, we reshape probabilistic map $P^i$ and perform a matrix multiplication between $P^i$ and $SM$. Finally, we obtain normalized $o^i_n = \sum_{m=1}^{N}({P^i}_mSM_{nm})$ and perform an element-wise sum operation with the probabilistic map $P^i$ to get the final output:

\begin{equation}
\label{inter-topo}
    {\hat{P}^i}_n = \alpha o^i_n + {P_n}^i 
\end{equation}
where $\alpha$ is initialized as 0 to capture stable probabilistic maps first. As training continues, we assign more weight on attention map so that the $\hat{P^i}$ at each position is a weighted sum across all positions and original probabilistic map $P^i$.

\subsection{Iterative Topology-Attention (ITA) Module}

\label{sec:iter}
As mentioned above, the critical points generated by persistent homology are sensitive, and consequently the attention map is also relatively unstable. 

By exploiting the correlation of probabilistic maps between different epochs, we can further improve the robustness of the obtained attention maps, which can lead to a better representation of the predicted probabilistic maps. Therefore, we introduce an ITA module to explore the relationships between the attention maps of different epochs. The iterative method not only helps with stability, but also enforces faster convergence. Formally, the ITA is calculated as $o_{T} = \beta o_{T-1} + (1 - \beta)o_{T}$.
Here, $\beta$ is a parameter to deal with the sensitiveness of the critical points, and $T$ denotes different training epochs. $o$ is the output of attention map which was described in Eq.~\ref{inter-topo}.
More details of ITA module is illustrated in Supplementary Fig. 1. During the training process, the final output $\hat{P}^i$ is generated by the sum of iterative attention map $o_T$ and the original probabilistic map $P^i$. Therefore, it has a global contextual view and selectively aggregates contexts according to the spatial attention map.

\section{Experiments}

We use three EM datasets with rich structure information to demonstrate the effectiveness of the proposed method. In this section, we will introduce the implementation details, datasets, and the experiment results on both datasets.

\myparagraph{Datasets.} We demonstrate the effectiveness of our proposed method with three different 3D Electron Microscopic Images datasets: \textbf{ISBI12}~\cite{arganda2015crowdsourcing}, \textbf{ISBI13}~\cite{arganda20133d} and \href{https://cremi.org/}{\textbf{CREMI}}. The size of \textbf{ISBI12}, \textbf{ISBI13} and \textbf{CREMI} are $30 \times 512 \times 512$, $100 \times 1024 \times 1024$ and $125 \times 1250 \times 1250$, respectively. 

\myparagraph{Train settings.}
We adopt ConvLSTM as our backbone architecture. Also, we apply simple data augmentation, Contrast-Limited Adaptive Histogram Equalization (CLAHE) and random flipping (for ISBI12 only to enlarge training size).
For the training parameters, we initialize learning rate ($lr$) as 0.001 and multiply by 0.5 every 50 epochs. We train our model with batch size of 15 for CREMI and ISBI13, and 30 for ISBI12. The number of training epochs are 35, 900 and 150 for CREMI, ISBI13 and ISBI12, respectively (without attention module). We use cross entropy loss as the optimization metric.

\myparagraph{Attention module details.} As described in Sec~\ref{sec:inter}, Topology-Attention module comes after ConvLSTM. We train the TACNet for another 15 epochs, with $lr=0.00001$. Specifically, the patch size is $39 \times 39$ for critical points extraction. The iterative rate $\beta$ is set as 0.5 for ITA module.

\begin{table}[t]
\fontsize{8}{6}

  \centering
    \caption{Experiment results for different models on CREMI dataset}

    \label{quan}
    
\begin{tabular}{c|c|c|c|c|c}
\hline\hline
\textsc {Datasets} & {Models} & {DICE} &{ARI} & {VOI} & {Betti Error}\\
\hline

\multirow{5}{*}{\textsc{CREMI}} &
DIVE &  0.9542 $\pm$ 0.0037 & 0.6532 $\pm$ 0.0247 & 2.513 $\pm$ 0.047  & 4.378 $\pm$ 0.152\\
~ & U-Net &  0.9523 $\pm$ 0.0049 & 0.6723 $\pm$ 0.0312 & 2.346 $\pm$ 0.105 & 3.016 $\pm$ 0.253\\
~ & Mosin. & 0.9489 $\pm$ 0.0053 & 0.7853 $\pm$ 0.0281 & 1.623 $\pm$ 0.083 & 1.973 $\pm$ 0.310\\
~ & TopoLoss &  0.9596 $\pm$ 0.0029 & 0.8083 $\pm$ 0.0104 & 1.462 $\pm$ 0.028 & 1.113 $\pm$ 0.224\\
~ & \textbf{TACNet} & \textbf{0.9665 $\pm$ 0.0008} & \textbf{0.8126 $\pm$ 0.0153} & \textbf{1.317 $\pm$ 0.165} &  \textbf{0.853 $\pm$ 0.183}\\
\hline

\multirow{5}{*}{\textsc{ISBI12}} &  
DIVE & 0.9709 $\pm$ 0.0029 & 0.9434 $\pm$ 0.0087 & 1.235 $\pm$ 0.025 & 3.187 $\pm$ 0.307\\

~ & U-Net &  0.9699 $\pm$ 0.0048 & 0.9338 $\pm$ 0.0072 & 1.367 $\pm$ 0.031 & 2.785 $\pm$ 0.269\\
~ & Mosin. & 0.9716 $\pm$ 0.0022 & 0.9312 $\pm$ 0.0052 & 0.983 $\pm$ 0.035 &1.238 $\pm$ 0.251\\
~ & TopoLoss & \textbf{0.9755 $\pm$ 0.0041} & \textbf{0.9444 $\pm$ 0.0076} & 0.782 $\pm$ 0.019  & 0.429 $\pm$ 0.104 \\
~ & \textbf{TACNet}  & 0.9576 $\pm$ 0.0047 & 0.9417 $\pm$ 0.0045 & \textbf{0.771 $\pm$ 0.027} &  \textbf{0.417 $\pm$ 0.117} \\
\hline

\multirow{5}{*}{\textsc{ISBI13}} &  
DIVE &  0.9658 $\pm$ 0.0020 & 0.6923 $\pm$ 0.0134 & 2.790 $\pm$ 0.025 & 3.875 $\pm$ 0.326 \\
~ & U-Net &  0.9649 $\pm$ 0.0057 & 0.7031 $\pm$ 0.0256 & 2.583 $\pm$ 0.078 & 3.463 $\pm$ 0.435\\
~ & Mosin. & 0.9623 $\pm$ 0.0047 & 0.7483 $\pm$ 0.0367 & 1.534 $\pm$ 0.063 & 2.952 $\pm$ 0.379\\
~ & TopoLoss & \textbf{0.9689 $\pm$ 0.0026} & \textbf{0.8064 $\pm$ 0.0112} & 1.436 $\pm$ 0.008 & 1.253 $\pm$ 0.172 \\
~ & \textbf{TACNet}  & 0.9510 $\pm$ 0.0022 & 0.7943 $\pm$ 0.0127 & \textbf{1.305 $\pm$ 0.016} & \textbf{1.175 $\pm$ 0.108} \\
\hline
\end{tabular}
\end{table}

\begin{figure}[tb!]
\centering 
\hspace{.0in}
\subfigure{
\includegraphics[width=0.12\textwidth]{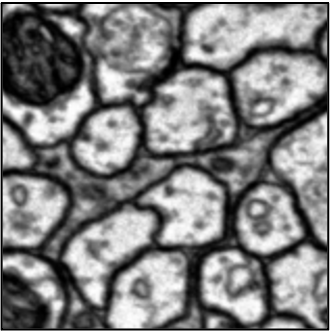}}
\hspace{-.08in}
\subfigure{
\includegraphics[width=0.12\textwidth]{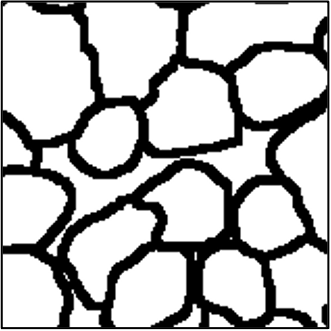}}
\hspace{-.08in}
\subfigure{
\includegraphics[width=0.12\textwidth]{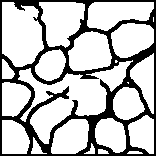}}
\hspace{-.08in}
\subfigure{
\includegraphics[width=0.12\textwidth]{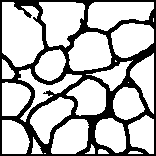}}
\hspace{-.08in}
\subfigure{
\includegraphics[width=0.12\textwidth]{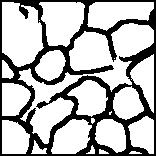}}
\hspace{-.08in}
\subfigure{
\includegraphics[width=0.12\textwidth]{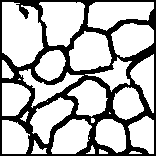}}
\vspace{-9pt}

\subfigure{
\includegraphics[width=0.12\textwidth]{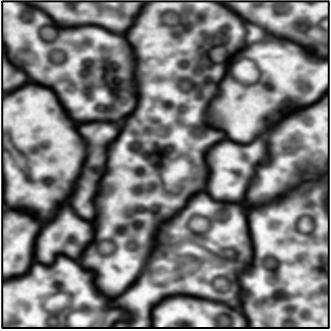}}
\hspace{-.08in}
\subfigure{
\includegraphics[width=0.12\textwidth]{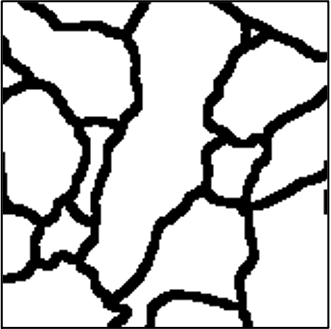}}
\hspace{-.08in}
\subfigure{
\includegraphics[width=0.12\textwidth]{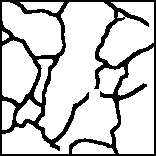}}
\hspace{-.08in}
\subfigure{
\includegraphics[width=0.12\textwidth]{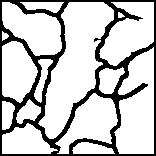}}
\hspace{-.08in}
\subfigure{
\includegraphics[width=0.12\textwidth]{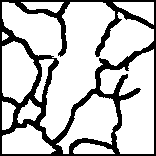}}
\hspace{-.08in}
\subfigure{
\includegraphics[width=0.12\textwidth]{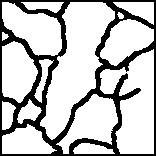}}

\vspace{-9pt}
\subfigure{
\includegraphics[width=0.12\textwidth]{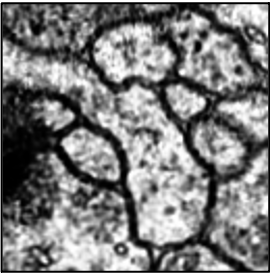}}
\hspace{-.08in}
\subfigure{
\includegraphics[width=0.12\textwidth]{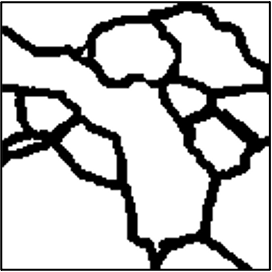}}
\hspace{-.08in}
\subfigure{
\includegraphics[width=0.12\textwidth]{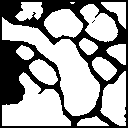}}
\hspace{-.08in}
\subfigure{
\includegraphics[width=0.12\textwidth]{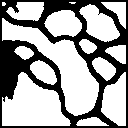}}
\hspace{-.08in}
\subfigure{
\includegraphics[width=0.12\textwidth]{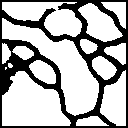}}
\hspace{-.08in}
\subfigure{
\includegraphics[width=0.12\textwidth]{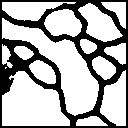}}

\caption{An illustration of structural accuracy. From left to right: a sample patch, the ground truth, results of UNet, TopoLoss, ConvLSTM and the proposed TACNet.}

\label{fig:illustration}

\end{figure}

\myparagraph{Quantitative and qualitative results.} In this paper we use similar topology-aware metrics as of~\cite{hu2019topology}, Adapted Rand Index (ARI), Variation of Information (VOI) and Betti number error. We also report dice scores for all the baselines and the proposed method. The details of the evaluation metrics can be found in the Sec.3 of~\cite{hu2019topology}. For all the experiments, we use three-fold cross-validation to report the average performance and standard deviation over the validation set. Tab.~\ref{quan} shows the quantitative results for the three different datasets. Note that we remove small connected components as a post-processing step to obtain final segmentation results. Our method generally outperforms existing methods~\cite{fakhry2016deep,ronneberger2015u,mosinska2018beyond} in terms of topology-aware metrics. Fig.~\ref{fig:illustration} shows qualitative results. Our method achieves better consistency/connection compared with other baselines.

In Tab.~\ref{quan}, the Betti Error of our TACNet brings $23.3\%$ improvement compared to the best baseline, TopoLoss~\cite{hu2019topology}, for CREMI dataset. Meanwhile, TACNet achieves best performances in terms of Betti Error and VOI on both ISBI12 and ISBI13 datasets. In Fig.~\ref{fig:illustration}, the results from TACNet possess better structures with less broken boundaries comparing to other baseline methods. Results show that our attention module strengthens the structural performance overall. Also, the proposed method computes topological information on a stack of 2D images rather than directly on a 3D image, and this significantly reduces the computational expense. Specifically, for CREMI dataset, our method takes $\approx$1.2 hours per epoch to train, whereas topoloss (3D version) takes $\approx$2.8h per epoch.

\myparagraph{Ablation study for TACNet.} Table.~\ref{ablation} shows the ablation study of the proposed method, which demonstrates the individual contributions of the two proposed modules, Spatial-Topology Attention and Iterative-Topology Attention. As shown in Table.~\ref{ablation}, compared with the backbone ConvLSTM model (Betti Error = 1.785), the STA improves the performance remarkably to 0.873. After applying the ITA module, the network further improves the performance by $2.3\%, 11.9\%, 2.1\%$ in Betti Error, VOI, and ARI, respectively. In addition, the combination of STA and ITA also improves the speed of convergence. For our ablation study, STA was trained with 50 epochs, but STA + ITA (our TACNet) was trained with fewer than 15 epochs for a better performance, which demonstrates that the ITA module can stabilize the training procedure.

\begin{table}[t]
\fontsize{8}{6}
  \centering
    \caption{Ablation study results for TACNet on CREMI dataset}
    \label{ablation}
    \begin{tabular}{c|c|c|c|c}
\hline\hline
\textsc{Models} & {DICE} &{ARI} & {VOI} & {Betti Error}\\
\hline
 
ConvLSTM  & 0.9667 $\pm$ 0.0007 & 0.7627 $\pm$ 0.0132  & 1.753 $\pm$ 0.212 & 1.785 $\pm$ 0.254 \\
ConvLSTM + STA  & 0.9663 $\pm$ 0.0004 & 0.7957 $\pm$ 0.0144 & 1.496 $\pm$ 0.156 & 0.873 $\pm$ 0.212 \\ 
Our TACNet & 0.9665 $\pm$ 0.0008 & \textbf{0.8126 $\pm$ 0.0153} & \textbf{1.317 $\pm$ 0.165} & \textbf{0.853 $\pm$ 0.183}   \\  
\hline
\end{tabular}
\end{table}

\myparagraph{Ablation study for number of input slices.} Table.~\ref{slices} is an illustration for the number of input slices. As shown in Table.~\ref{slices}, compared with 1 slice (Betti Error = 1.386) or 5 slices (Betti Error = 0.967), the adopted setting of 3 slices achieves the best results (Betti Error = 0.853). It's not surprised that the 3 slices achieves better performance than 1 slice, as it makes use of inter-slice information. On the other hand, the dataset is anisotropic, and the slices further away are increasingly different from the center slice, which degrades the performance for 5 slices setting.

\begin{table}[t]
  \centering
    \caption{Ablation study results for number of input slices on CREMI}
    \label{slices}
    \begin{tabular}{c|c|c|c|c}
\hline\hline
\textsc {Number} & {ARI} & {VOI} & {Betti Error} & {Time}\\
\hline

1s  & 0.7813 $\pm$ 0.0141  & 1.672 $\pm$ 0.191 & 1.386 $\pm$ 0.117 & 0.99h/epoch \\
3s & \textbf{0.8126 $\pm$ 0.0153} & \textbf{1.317 $\pm$ 0.165} & \textbf{0.853 $\pm$ 0.183} & 1.20h/epoch\\ 
5s & {0.8076 $\pm$ 0.0107} & {1.461 $\pm$ 0.125} & {0.967 $\pm$ 0.098} & 2.78h/epoch \\ 
\hline
\end{tabular}
\end{table}

\begin{figure}[tb!]
\centering 
\subfigure{
\includegraphics[width=0.15\textwidth]{ablation/46org.png}}
\hspace{-.08in}
\subfigure{
\includegraphics[width=0.15\textwidth]{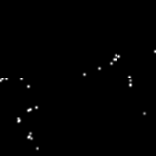}}
\hspace{-.08in}
\subfigure{
\includegraphics[width=0.15\textwidth]{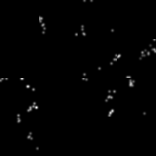}}
\hspace{-.08in}
\subfigure{
\includegraphics[width=0.15\textwidth]{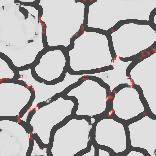}}

\vspace{-9pt}
\subfigure{
\includegraphics[width=0.15\textwidth]{ablation/59org.png}}
\hspace{-.08in}
\subfigure{
\includegraphics[width=0.15\textwidth]{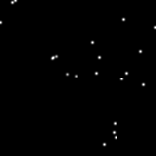}}
\hspace{-.08in}
\subfigure{
\includegraphics[width=0.15\textwidth]{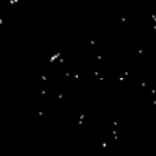}}
\hspace{-.08in}
\subfigure{
\includegraphics[width=0.15\textwidth]{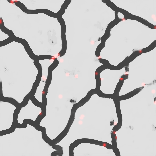}}

\caption{Illustration of the proposed Topology-Attention. From left to right: original images $S^i$, the smooth critical points of $CP^i$, final attention map $o^i$, and the final probability map $\hat{P}^i$ with $o^i$ superimposed in red (Zoom in and best viewed in color).}

\label{fig:attention}

\end{figure}

\myparagraph{Illustration of the attention module.}
We select two images to show the effectiveness of the attention module in Fig.~\ref{fig:attention}. Compared with the second column showing only the critical points detected in the current slice, the attention map in the third column captures more information with the structural similarity from adjacent slices. The attention areas on the final probabilistic map (last column) are highlighted with red color. We observe that the responses of most broken connections are high with attention module enhancement. In summary, Fig.~\ref{fig:attention} demonstrates that our TACNet successfully captures the structure information and further improves responses on those essential areas.

\section{Conclusion}

In this paper, we proposed a novel Topology Attention Module with ConvLSTM, named TACNet, for 3D EM image segmentation. The Topology-Attention module includes spatial and iterative components, which capture rich structural context and stabilize the attention mechanism. The illustration of the attention map shows that our topology attention module can identify the regions of broken structures, and generates more complete structural results. Validated with three EM anisotropic datasets, our method outperforms baselines in terms of topology-aware metrics. We expect the performance to further improve on isotropic datasets, because slices are closer (due to higher sampling rate in the z-dimension) with more consistent topologies across slices. For the future work, we will apply TACNet to datasets of other medical structures, such as cardiac and vascular images, to prove its efficacy in a broader medical domain.
\bibliographystyle{splncs04}
\bibliography{ref}

\end{document}


\title{Topology-Attention ConvLSTM Network for 3D Image Segmentation \\ --- Supplementary Material ---}
\author{Jiaqi Yang\inst{1\star}\and
Xiaoling Hu\inst{2\star} \and
Chao Chen\inst{2} \and
Chialing Tsai\inst{1} }

\renewcommand{\thefootnote}{\fnsymbol{footnote}}
\footnotetext[1]{The two authors contributed equally to this paper.}

\institute{Graduate Center, CUNY \and Stony Brook University}
\maketitle              
\section{Illustration of Iterative Topology-Attention module}
\begin{figure}[th!]
\centering
\includegraphics[width=0.9\columnwidth]{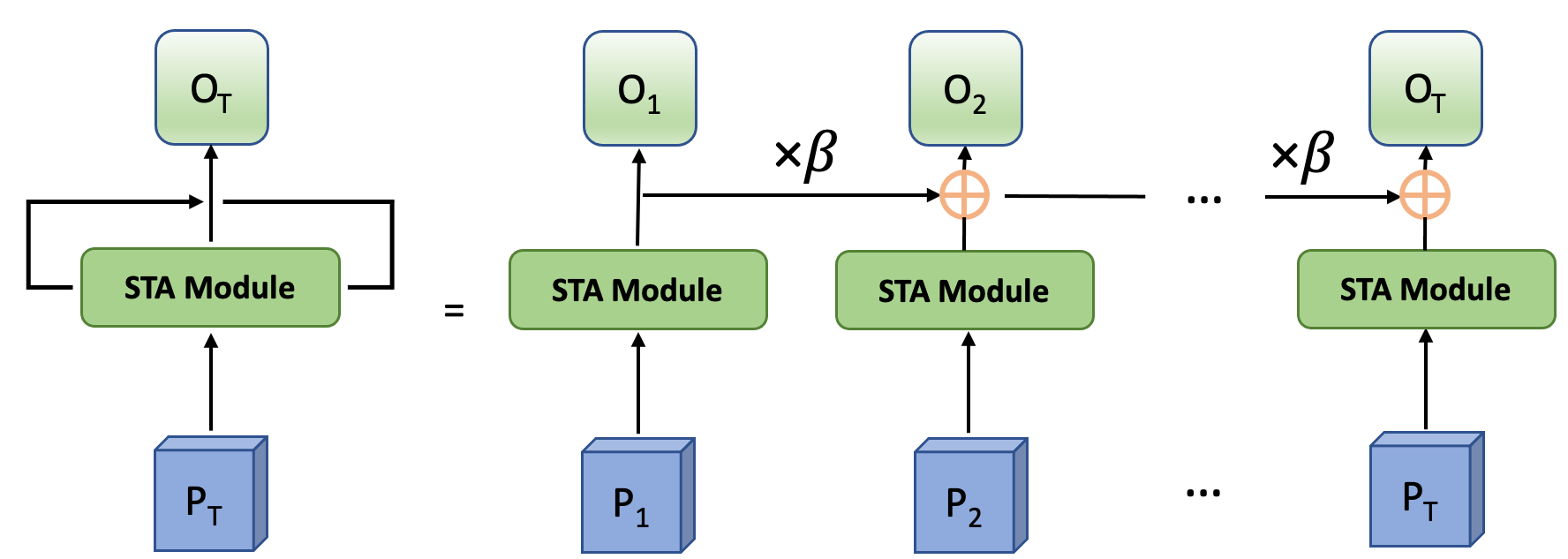} 
\caption{Illustration of Iterative Topology-Attention (ITA) module. $p_T$ and $o_{T}$ are the probabilistic map and the output of attention map at $T$ epoch. }
\label{temporal}
\end{figure}

\bibliographystyle{splncs04}
\bibliography{ref}